\documentclass[copyright]{eptcs}
\providecommand{\event}{PLACES 2017} 
\usepackage{breakurl}                
\usepackage{underscore}              

\usepackage{listings,xspace}
\usepackage{xcolor}
\usepackage{graphicx}
\usepackage{url}

\usepackage{bold-extra}

\lstdefinelanguage{Scala}%
{morekeywords={abstract,case,catch,char,class,%
    def,else,extends,final,%
    if,import,%
    match,module,new,null,object,override,package,private,protected,%
    public,return,super,this,throw,trait,try,type,val,var,with,implicit,%
    macro,sealed,%
  },%
  sensitive,%
  morecomment=[l]//,%
  morecomment=[s]{/*}{*/},%
  morestring=[b]",%
  morestring=[b]',%
  showstringspaces=false%
}[keywords,comments,strings]%

\newcommand{\comment}[1]{}

\newcommand{\eg}{{\em e.g.,~}}

\title{Quantifying and Explaining Immutability in Scala}
\author{Philipp Haller
\institute{KTH Royal Institute of Technology\\ Stockholm, Sweden}
\email{phaller@kth.se}
\and
Ludvig Axelsson
\institute{KTH Royal Institute of Technology\\ Stockholm, Sweden}
\email{ludvigax@kth.se}
}
\def\titlerunning{Quantifying and Explaining Immutability in Scala}
\def\authorrunning{P. Haller \& L. Axelsson}

\begin{document}
\maketitle

\begin{abstract}
Functional programming typically emphasizes programming with first-class functions and immutable data. Immutable data types enable fault tolerance in distributed systems, and ensure process isolation in message-passing concurrency, among other applications. However, beyond the distinction between reassignable and non-reassignable fields, Scala's type system does not have a built-in notion of immutability for type definitions. As a result, immutability is ``by-convention'' in Scala, and statistics about the use of immutability in real-world Scala code are non-existent.

This paper reports on the results of an empirical study on the use of immutability in several medium-to-large Scala open-source code bases, including Scala's standard library and the Akka actor framework. The study investigates both shallow and deep immutability, two widely-used forms of immutability in Scala. Perhaps most interestingly, for type definitions determined to be mutable, explanations are provided for why neither the shallow nor the deep immutability property holds; in turn, these explanations are aggregated into statistics in order to determine the most common reasons for why type definitions are mutable rather than immutable.
\end{abstract}

\section{Introduction}

Immutability is an important property of data types, especially in the
context of concurrent and distributed programming. For example,
objects of immutable type may be safely shared by concurrent processes
without the possibility of data races. In message-passing concurrency,
sending immutable messages helps ensure process isolation. Finally, in
distributed systems immutability enables efficient techniques for
providing fault tolerance.

Scala's type system does not have a built-in notion of immutability
for type definitions. Instead, immutability is "by-convention" in
Scala. In addition, statistics about the use of immutability in
real-world Scala code are non-existent. This is problematic, since
such statistics could inform extensions of Scala's type system for
enforcing immutability properties.

\paragraph{Contributions}
This paper presents the first empirical results evaluating the
prevalence of immutability in medium-to-large open-source Scala code
bases, including the Scala standard library and the Akka actor
framework~\cite{Akka}. We considered three different immutability
properties, all of which occur frequently in all our case studies.  In
addition, we provide empirical results, evaluating causes for
mutability of type definitions.

\section{Immutability Analysis}\label{sec:analysis}

This paper uses a notion of immutability that applies to {\em type
  definitions} rather than object references as in other
work~\cite{TschantzE05,GordonPPBD12}. For example, the definition of
an immutable class implies that all its instances are immutable.  We
refer to class, trait, and object definitions collectively as {\em
  templates}, following the terminology of the Scala language
specification~\cite{Odersky14}.

We distinguish three different immutability properties: (a) deep
immutability, (b) shallow immutability, and (c) conditional deep
immutability. Deep immutability is the strongest property; it requires
that none of the declared or inherited fields is reassignable, and
that the types of all declared or inherited fields are deeply
immutable. Shallow immutability requires that none of the parents is
mutable and that none of the declared or inherited fields is
reassignable. Conditional deep immutability requires that none of the
declared or inherited fields is reassignable, and that the types of
all declared or inherited fields are deeply immutable, unless they are
abstract types. For example, the type parameter \verb|T| of the
generic class \verb|Option[T]| is abstract; type \verb|T| is unknown
within the definition of type \verb|Option[T]|. Similarly, a Scala
abstract type member~\cite{AminGORS16} is treated as an abstract
type. Finally, a class that declares or inherits a reassignable field
(a Scala \verb|var|) is {\em mutable}.

\subsection{Implementation}

We implement our analysis as a compiler plugin for Scala
2.11.x.\footnote{See
  \url{https://github.com/luax/scala-immutability-plugin}} The plugin
can be enabled when building Scala projects using the sbt or Maven
build tools. The immutability analysis is implemented using Reactive
Async~\cite{Haller16} which extends LVars~\cite{KuperTKN14},
lattice-based variables, with cyclic dependency resolution. For each
template definition we maintain a ``cell'' that keeps track of the
immutability property of the template. The value of the cell is taken
from an immutability lattice; the analysis may update cell values
monotonically according to the immutability lattice, based on evidence
found during the analysis. For example, the cell value of a subclass
is updated to \verb|Mutable| when the analysis detects that one of the
superclasses is mutable. Initially, all templates are assumed to be
deeply immutable; this assumption is then updated incrementally based
on evidence found by the analysis.

\section{Empirical Study}\label{sec:results}

We evaluate the prevalence of the immutability properties defined in
Section~\ref{sec:analysis} in four medium-to-large Scala open-source
projects: Scala's standard library (version 2.11.8), Akka's actor
package (version 2.4.17), ScalaTest (version 3.0.1), and
Signal/Collect (version 8.0.2).

The Scala standard library consists of 33107 source lines of code
(excluding blank lines and comments).\footnote{Measured using cloc
  v1.70, see \url{https://github.com/AlDanial/cloc}} The library
includes an extensive collection package~\cite{OderskyM09} with both
mutable and immutable collection types, as well as math, I/O, and
concurrency packages such as futures~\cite{Futures}. Certain packages are designed to
only define immutable types, including package
\verb|scala.collection.immutable| and package
\verb|scala.collection.parallel.immutable|. Other packages are
designed to define mutable types, including packages
\verb|scala.collection.mutable|, \verb|scala.collection.concurrent|,
and \verb|scala.collection.parallel.mutable|.

Akka's actor package is the standard actor implementation for
Scala. ScalaTest~\cite{ScalaTest} is the most widely-used testing
framework for Scala. Signal/Collect~\cite{StutzBC10} is a distributed
graph processing framework based on Akka.

Our empirical study aims to answer the following two main research
questions:
\begin{itemize}
  \item[RQ1] How frequent is each immutability property for classes,
    traits, and objects?
  \item[RQ2] For classes/traits/objects that are not deeply immutable:
    what are the most common reasons why stronger immutability
    properties are not satisfied?
\end{itemize}

\begin{table*}
  \begin{tabular}{ | l | r | r | r | r | r | }
  \hline
  \textbf{Template} & \multicolumn{1}{l|}{\textbf{Occurrences}} & \multicolumn{1}{l|}{\textbf{Mutable}} & \multicolumn{1}{l|}{\textbf{Shallow}} & \multicolumn{1}{l|}{\textbf{Deep}} & \multicolumn{1}{l|}{\textbf{Cond. Deep}} \\
  \hline
    Class             & 626 (33,5\%)                              & 330 (52,7\%)                          & 54 (8,6\%)                            & 124 (19,8\%)                       & 118 (18,8\%)                                     \\ \hline
    Case class        & 75 (4,0\%)                                & 19 (25,3\%)                           & 7 (9,3\%)                             & 9 (12,0\%)                         & 40 (53,3\%)                                      \\ \hline
    Anon. class   & 330 (17,7\%)                              & 209 (63,3\%)                          & 26 (7,9\%)                            & 95 (28,8\%)                        & 0 (0\%)                                          \\ \hline
    Trait             & 466 (25,0\%)                              & 224 (48,1\%)                          & 15 (3,2\%)                            & 93 (20,0\%)                        & 134 (28,8\%)                                     \\ \hline
    Object            & 358 (19,2\%)                              & 106 (29,6\%)                          & 29 (8,1\%)                            & 223 (62,3\%)                       & 0 (0\%)                                          \\ \hline
    Case object       & 12 (0,6\%)                                & 3 (25,0\%)                            & 0 (0\%)                               & 9 (75,0\%)                         & 0 (0\%)                                          \\ \hline
    \textbf{Total}    & \textbf{1867 (100,0\%)}                   & \textbf{891 (47,7\%)}                 & \textbf{131 (7,0\%)}                  & \textbf{553 (29,6\%)}              & \textbf{292 (15,6\%)}                            \\ \hline
  \end{tabular}
  \caption{Immutability statistics for Scala standard library.}\label{tab:results}
\end{table*}

\subsection{Research Question 1}

Tables~\ref{tab:results} shows the immutability statistics for Scala's
standard library. One of the most important results is that {\em the
  majority of classes/traits/objects in Scala's standard library
  satisfy one of the immutability properties.} This confirms the
intuition that functional programming with immutable types is an
important programming style in Scala. Interestingly, the most common
immutability property for case classes and traits is conditional deep
immutability. Thus, {\em whether a case class or trait is deeply
  immutable in most cases depends on the instantiation of type
  parameters or abstract types.}  In contrast, the majority of classes
that are not case classes is mutable. Note that objects and anonymous
classes cannot be conditionally deeply immutable, since these
templates cannot have type parameters or abstract type members.

\begin{table*}
  \begin{tabular}{ | l | r | r | r | r | r | }
  \hline
    \textbf{Template} & \multicolumn{1}{l|}{\textbf{Occurrences}} & \multicolumn{1}{l|}{\textbf{Mutable}} & \multicolumn{1}{l|}{\textbf{Shallow}} & \multicolumn{1}{l|}{\textbf{Deep}} & \multicolumn{1}{l|}{\textbf{Cond. Deep}} \\ \hline
    Class             & 299 (26,8\%)                              & 115 (38,5\%)                          & 93 (31,1\%)                           & 82 (27,4\%)                        & 9 (3,0\%)                                        \\ \hline
    Case class        & 206 (18,4\%)                              & 23 (11,2\%)                           & 64 (31,1\%)                           & 90 (43,7\%)                        & 29 (14,1\%)                                      \\ \hline
    Anon. class   & 77 (6,9\%)                                & 33 (42,9\%)                           & 8 (10,4\%)                            & 36 (46,8\%)                        & 0 (0\%)                                          \\ \hline
    Trait             & 239 (21,4\%)                              & 22 (9,2\%)                            & 17 (7,1\%)                            & 140 (58,6\%)                       & 60 (25,1\%)                                      \\ \hline
    Object            & 220 (19,7\%)                              & 9 (4,1\%)                             & 47 (21,4\%)                           & 164 (74,5\%)                       & 0 (0\%)                                          \\ \hline
    Case object       & 76 (6,8\%)                                & 2 (2,6\%)                             & 0 (0\%)                               & 74 (97,4\%)                        & 0 (0\%)                                          \\ \hline
    \textbf{Total}    & \textbf{1117 (100,0\%)}                   & \textbf{204 (18,3\%)}                 & \textbf{229 (20,5\%)}                 & \textbf{586 (52,5\%)}              & \textbf{98 (8,8\%)}                              \\ \hline
  \end{tabular}
  \caption{Immutability statistics for Akka (\texttt{akka-actor} package). }\label{tab:results-akka}
\end{table*}

Table~\ref{tab:results-akka} shows the immutability statistics for
Akka. The percentage of mutable classes/traits/objects is
significantly lower compared to Scala's standard library (18.3\% for
Akka versus 47.7\% for the Scala library).

\begin{table*}
  \begin{tabular}{ | l | r | r | r | r | r | }
  \hline
    \textbf{Template} & \multicolumn{1}{l|}{\textbf{Occurrences}} & \multicolumn{1}{l|}{\textbf{Mutable}} & \multicolumn{1}{l|}{\textbf{Shallow}} & \multicolumn{1}{l|}{\textbf{Deep}} & \multicolumn{1}{l|}{\textbf{Cond. Deep}} \\ \hline
    Class             & 791 (36,1\%)                              & 216 (27,3\%)                          & 249 (31,5\%)                          & 288 (36,4\%)                       & 38 (4,8\%)                                       \\ \hline
    Case class        & 153 (7,0\%)                               & 15 (9,8\%)                            & 81 (52,9\%)                           & 54 (35,3\%)                        & 3 (2,0\%)                                        \\ \hline
    Anon. class   & 688 (31,4\%)                              & 200 (29,1\%)                          & 293 (42,6\%)                          & 195 (28,3\%)                       & 0 (0\%)                                          \\ \hline
    Trait             & 227 (10,3\%)                              & 61 (26,9\%)                           & 45 (19,8\%)                           & 91 (40,1\%)                        & 30 (13,2\%)                                      \\ \hline
    Object            & 254 (11,6\%)                              & 19 (7,5\%)                            & 18 (7,1\%)                            & 217 (85,4\%)                       & 0 (0\%)                                          \\ \hline
    Case object       & 81 (3,7\%)                                & 2 (2,5\%)                             & 0 (0\%)                               & 79 (97,5\%)                        & 0 (0\%)                                          \\ \hline
    \textbf{Total}    & \textbf{2194 (100,0\%)}                   & \textbf{513 (23,4\%)}                 & \textbf{686 (31,3\%)}                 & \textbf{924 (42,1\%)}              & \textbf{71 (3,2\%)}                              \\ \hline
  \end{tabular}
  \caption{Immutability statistics for ScalaTest.}\label{tab:results-scalatest}
\end{table*}

\begin{table*}
  \begin{tabular}{ | l | r | r | r | r | r | }
  \hline
    \textbf{Template} & \multicolumn{1}{l|}{\textbf{Occurrences}} & \multicolumn{1}{l|}{\textbf{Mutable}} & \multicolumn{1}{l|}{\textbf{Shallow}} & \multicolumn{1}{l|}{\textbf{Deep}} & \multicolumn{1}{l|}{\textbf{Cond. Deep}} \\ \hline
    Class             & 160 (58,0\%)                              & 78 (48,8\%)                           & 24 (15,0\%)                           & 14 (8,8\%)                         & 44 (27,5\%)                                      \\ \hline
    Case class        & 42 (15,2\%)                               & 4 (9,5\%)                             & 11 (26,2\%)                           & 15 (35,7\%)                        & 12 (28,6\%)                                      \\ \hline
    Anon. class   & 4 (1,4\%)                                 & 4 (100,0\%)                           & 0 (0\%)                               & 0 (0\%)                            & 0 (0\%)                                          \\ \hline
    Trait             & 24 (8,7\%)                                & 6 (25,0\%)                            & 1 (4,2\%)                             & 3 (12,5\%)                         & 14 (58,3\%)                                      \\ \hline
    Object            & 41 (14,9\%)                               & 19 (46,3\%)                           & 5 (12,2\%)                            & 17 (41,5\%)                        & 0 (0\%)                                          \\ \hline
    Case object       & 5 (1,8\%)                                 & 0 (0\%)                               & 0 (0\%)                               & 5 (100,0\%)                        & 0 (0\%)                                          \\ \hline
    \textbf{Total}    & \textbf{276 (100,0\%)}                    & \textbf{111 (40,2\%)}                 & \textbf{41 (14,9\%)}                  & \textbf{54 (19,6\%)}               & \textbf{70 (25,4\%)}                             \\ \hline
  \end{tabular}
  \caption{Immutability statistics for Signal/Collect.}\label{tab:results-signal}
\end{table*}

Table~\ref{tab:results-signal} shows the immutability statistics for
Signal/Collect. Unique to Signal/Collect is the high percentage of
mutable singleton objects (46.3\%), which ranges between 4.1\% (Akka)
and 29.6\% (Scala library). However, also in Signal/Collect is the
percentage of mutable case classes low compared to other kinds of
templates.

\paragraph{Summary}

In our case studies, the majority of classes/traits/objects satisfy
one of our immutability properties. The prevalence of mutability is
especially low for case classes (with structural equality) and
singleton objects. Except for Signal/Collect, which is unique in this
case, the majority of singleton objects are deeply immutable, ranging
between 62.3\% and 85.4\% in our case studies. The percentage of
deeply immutable {\em case objects} is even higher, ranging between
75\% and 100\%, including Signal/Collect.

\begin{table*}[t!]
  \begin{tabular}{ | l | l | l | }
  \hline
  {\bfseries ~Reason~} & {\bfseries ~Immutability Property~} & {\bfseries ~Attribute Key~} \\
  \hline
  Parent type mutable (assumption)                & Mutable           & A \\
  \hline
  Parent type mutable                             & Mutable           & B \\
  \hline
  Reassignable field (public)                     & Mutable           & C \\
  \hline
  Reassignable field (private)                    & Mutable           & D \\
  \hline
  Parent type unknown                             & Mutable           & E \\
  \hline
  Parent type shallow immutable                   & Shallow immutable & F \\
  \hline
  \verb|val| field with unknown type              & Shallow immutable & G \\
  \hline
  \verb|val| field with mutable type              & Shallow immutable & H \\
  \hline
  \verb|val| field with mutable type (assumption) & Shallow immutable & I \\
  \hline
  \end{tabular}
  \caption{Template attributes and their influence on immutability
    properties.}\label{tab:properties}
\end{table*}

In order to answer RQ2, we identified nine template {\em attributes},
shown in Table~\ref{tab:properties}, which explain why certain
immutability properties cannot be satisfied. The presence of the first
five attributes forces the corresponding template to be classified as
mutable. For example, a template is classified as mutable if it
declares a reassignable field (attributes C and D). The last four
attributes prevent the corresponding template from satisfying either
deep of conditionally deep immutability. For example, if a parent
class or trait is only shallow immutable (but not deeply immutable),
then the corresponding template cannot be deeply immutable or
conditionally deeply immutable either (attribute F).

\subsection{Research Question 2}

\begin{table}
\begin{minipage}[b]{0.45\linewidth}
    \begin{tabular}{|l|r|}
        \hline
        \textbf{Attribute(s)} & \multicolumn{1}{l|}{\textbf{Occurrences}} \\ \hline
        B                     & 609 (68,4\%)                              \\ \hline
        B C                   & 71 (8,0\%)                                \\ \hline
        B C D                 & 1 (0,1\%)                                 \\ \hline
        B D                   & 19 (2,1\%)                                \\ \hline
        B E                   & 7 (0,8\%)                                 \\ \hline
        C                     & 26 (2,9\%)                                \\ \hline
        C D                   & 1 (0,1\%)                                 \\ \hline
        D                     & 87 (9,8\%)                                \\ \hline
        D E                   & 4 (0,4\%)                                 \\ \hline
        E                     & 66 (7,4\%)                                \\ \hline
    \end{tabular}
    \caption{Scala library: attributes causing mutability.}\label{tab:mutable-scalalib}
\end{minipage}
\begin{minipage}[b]{0.45\linewidth}
    \begin{tabular}{|l|r|}
        \hline
        \textbf{Attribute(s)} & \multicolumn{1}{l|}{\textbf{Occurrences}} \\ \hline
        F                     & 28 (21,4\%)                               \\ \hline
        F G                   & 5 (3,8\%)                                 \\ \hline
        F G H                 & 1 (0,8\%)                                 \\ \hline
        F H                   & 4 (3,1\%)                                 \\ \hline
        F J                   & 6 (4,6\%)                                 \\ \hline
        G                     & 22 (16,8\%)                               \\ \hline
        G H                   & 4 (3,1\%)                                 \\ \hline
        G H J                 & 3 (2,3\%)                                 \\ \hline
        G J                   & 2 (1,5\%)                                 \\ \hline
        H                     & 40 (30,5\%)                               \\ \hline
        H J                   & 3 (2,3\%)                                 \\ \hline
        J                     & 7 (5,3\%)                                 \\ \hline
    \end{tabular}
    \caption{Scala library: attributes causing shallow immutability (instead of deep immutability).}\label{tab:shallow-scalalib}
\end{minipage}
\end{table}

Tables~\ref{tab:mutable-scalalib}~and~\ref{tab:shallow-scalalib} show
the causes for mutability and shallow immutability, respectively, for
the Scala library. The main cause for a template to be classified as
mutable is the existence of a parent which is mutable. Important
causes for templates to be classified as shallow immutable rather than
deeply immutable are (a) the existence of a non-reassignable field
with a mutable type (attribute H), and (b) the existence of a parent
which is shallow immutable (attribute F).

\begin{table}
\centering
\begin{minipage}[b]{0.45\linewidth}
    \begin{tabular}{|l|r|}
        \hline
        \textbf{Attribute(s)} & \multicolumn{1}{l|}{\textbf{Occurrences}} \\ \hline
        A                     & 3 (1,5\%)                                   \\ \hline
        A B D                 & 1 (0,5\%)                                   \\ \hline
        A E                   & 1 (0,5\%)                                   \\ \hline
        B                     & 76 (37,3\%)                                 \\ \hline
        B C                   & 3 (1,5\%)                                   \\ \hline
        B C D                 & 1 (0,5\%)                                   \\ \hline
        B D                   & 6 (2,9\%)                                   \\ \hline
        B E                   & 6 (2,9\%)                                   \\ \hline
        C                     & 7 (3,4\%)                                   \\ \hline
        C D                   & 2 (1,0\%)                                   \\ \hline
        C D E                 & 1 (0,5\%)                                   \\ \hline
        D                     & 24 (11,8\%)                                 \\ \hline
        D E                   & 1 (0,5\%)                                   \\ \hline
        E                     & 72 (35,3\%)                                 \\ \hline
    \end{tabular}
    \caption{Akka: attributes causing mutability.}\label{tab:mutable-akka}
\end{minipage}
\begin{minipage}[b]{0.45\linewidth}
    \begin{tabular}{|l|r|}
        \hline
        \textbf{Attribute(s)} & \multicolumn{1}{l|}{\textbf{Occurrences}} \\ \hline
        F                     & 38 (16,6\%)                               \\ \hline
        F G                   & 9 (3,9\%)                                 \\ \hline
        F G H                 & 2 (0,9\%)                                 \\ \hline
        F G J                 & 3 (1,3\%)                                 \\ \hline
        F H                   & 3 (1,3\%)                                 \\ \hline
        F J                   & 3 (1,3\%)                                 \\ \hline
        G                     & 94 (41,0\%)                               \\ \hline
        G H                   & 8 (3,5\%)                                 \\ \hline
        G H I                 & 1 (0,4\%)                                 \\ \hline
        G H J                 & 1 (0,4\%)                                 \\ \hline
        G J                   & 16 (7,0\%)                                \\ \hline
        H                     & 22 (9,6\%)                                \\ \hline
        H J                   & 4 (1,7\%)                                 \\ \hline
        J                     & 25 (10,9\%)                               \\ \hline
    \end{tabular}
    \caption{Akka: attributes causing shallow immutability (instead of deep immutability).}\label{tab:shallow-akka}
\end{minipage}
\end{table}

Tables~\ref{tab:mutable-akka}~and~\ref{tab:shallow-akka} show the
causes for mutability and shallow immutability, respectively, for Akka
actors. The main cause for a template to be classified as mutable is
the existence of a parent which is mutable; this matches the
statistics of the Scala library. Other important causes are (a) parent
types whose immutability is unknown (\eg due to third-party libraries
for which no analysis results are available) and (b) private
reassignable fields. Unlike the Scala library, the most important
cause for shallow immutability (rather than deep immutability) in Akka
are non-reassignable fields of a type whose immutability is unknown;
this suggests that the absence of analysis results for third-party
libraries has a significant impact on the classification of a type as
shallow immutable rather than deeply immutable. On the other hand,
this means that the actual percentage of deeply immutable templates
may be even higher. Therefore, an important avenue for future work is
to enable the analysis of third-party libraries. The second most
important cause is the existence of a parent which is shallow
immutable (attribute F).

\section{Conclusion}\label{sec:conclusion}

Immutability is an important property of data types, especially in the
context of concurrent and distributed programming. For example,
objects of immutable type may be safely shared by concurrent processes
without the possibility of data races. In message-passing concurrency,
sending immutable messages helps ensure process isolation. In this
paper we presented the first empirical results evaluating the
prevalence of immutability in medium-to-large open-source Scala code
bases, including the Scala standard library and the Akka actor
framework. We considered three different immutability properties, all
of which occur frequently in all our case studies. In our case
studies, the majority of classes/traits/objects satisfy one of our
immutability properties. The prevalence of mutability is especially
low for case classes (classes with structural equality) and singleton
objects. The most important causes for mutability are mutable parent
classes and private reassignable fields. To our knowledge we presented
the first empirical study of its kind. We believe our insights are
valuable both for informing the further evolution of the Scala
language, and for designers of new wide-spectrum languages, combining
functional and imperative features.

\bibliographystyle{eptcs}
\bibliography{bibliography}
\end{document}